\documentclass[sigconf]{acmart}

\AtBeginDocument{%
  \providecommand\BibTeX{{%
    Bib\TeX}}}

\usepackage{enumitem}

\usepackage{amsfonts,multirow, mathtools,bm}
\usepackage{stfloats}
\usepackage{algorithmic}

\usepackage{braket}

\def\BibTeX{{\rm B\kern-.05em{\sc i\kern-.025em b}\kern-.08em
    T\kern-.1667em\lower.7ex\hbox{E}\kern-.125emX}}
    
\setcopyright{none} 

\begin{document}

\twocolumn

\title{MUSS-TI: Multi-level Shuttle Scheduling for Large-Scale Entanglement Module Linked Trapped-Ion}

\author{Xian Wu}
\authornotemark[1]
\affiliation{%
  \institution{The Hong Kong University of Science and Technology (Guangzhou)}
  \city{Guangzhou}
  \country{China}
  \postcode{511453}
}
\email{xwu574@connect.hkust-gz.edu.cn}
\author{Chenghong Zhu}
\authornote{Co-first authors.}
\affiliation{%
  \institution{The Hong Kong University of Science and Technology (Guangzhou)}
  \city{Guangzhou}
  \country{China}
  \postcode{511453}
}
\email{czhu854@connect.hkust-gz.edu.cn}

\author{Jingbo Wang}
\authornote{Co-corresponding authors.}
\affiliation{\institution{Beijing Academy of Quantum Information Sciences}
\city{Beijing}
\country{China}
}
\email{wangjb@baqis.ac.cn}

\author{Xin Wang}
\authornotemark[2]
\affiliation{%
  \institution{The Hong Kong University of Science and Technology (Guangzhou)}
  \city{Guangzhou}
  \country{China}
  \postcode{511453}
}
\email{felixxinwang@hkust-gz.edu.cn}


\begin{abstract}

Trapped-ion computing is a leading architecture in the pursuit of scalable and high fidelity quantum systems. Modular quantum architectures based on photonic interconnects offer a promising path for scaling trapped ion devices. In this design, multiple Quantum Charge Coupled Device (QCCD) units are interconnected through entanglement module. Each unit features a multi-zone layout that separates functionalities into distinct areas, enabling more efficient and flexible quantum operations. However, achieving efficient and scalable compilation of quantum circuits in such entanglement module linked Quantum Charge-Coupled Device (EML-QCCD) remains a primary challenge for practical quantum applications. 

In this work, we propose a scalable compiler tailored for large-scale trapped-ion architectures, with the goal of reducing the shuttling overhead inherent in EML-QCCD devices. MUSS-TI introduces a multi-level scheduling approach inspired by multi-level memory scheduling in classical computing. This method is designed to be aware of the distinct roles of different zones and to minimize the number of shuttling operations required in EML-QCCD systems. We demonstrate that EML-QCCD architectures are well-suited for executing large-scale applications. Our evaluation shows that MUSS-TI reduces shuttle operations by 41.74\% for applications with 30–32 qubits, and by an average of 73.38\% and 59.82\% for applications with 117–128 qubits and 256–299 qubits, respectively. 

\end{abstract}




\maketitle

\section{Introduction}
Quantum computing holds immense promise for humanity, offering solutions to problems that are currently intractable for classical computers, such as simulating complex quantum systems, optimizing logistical operations, and solving cryptographic challenges. Trapped-ion quantum computing has emerged as one of the most promising architectures for realizing large-scale quantum computation due to its ability to perform high-fidelity gate operations, long coherence times, and the inherent scalability of trapped ion systems~\cite{bruzewicz2019trapped, haffner2008quantum}. Recent developments in quantum error correction, advanced laser control, and photonic interconnects have enabled remarkable progress in the implementation of quantum algorithms using trapped ions~\cite{zhang2020error, hilder2022fault}. 

Despite the many advantages of trapped-ion quantum computing, scalability remains a key challenge.
To address the limitation on the number of qubits that can be reliably controlled within a single trap, the QCCD (Quantum Charge-Coupled Device) offers a scalable solution.
QCCD architecture leverages shuttle-based quantum operations to interconnect multiple ion-trap zones, providing a practical pathway for scaling up trapped-ion qubit systems~\cite{kielpinski2002architecture, pino2021demonstration}.

\begin{figure}[t]
    \centering
    \includegraphics[width=1.0\linewidth]{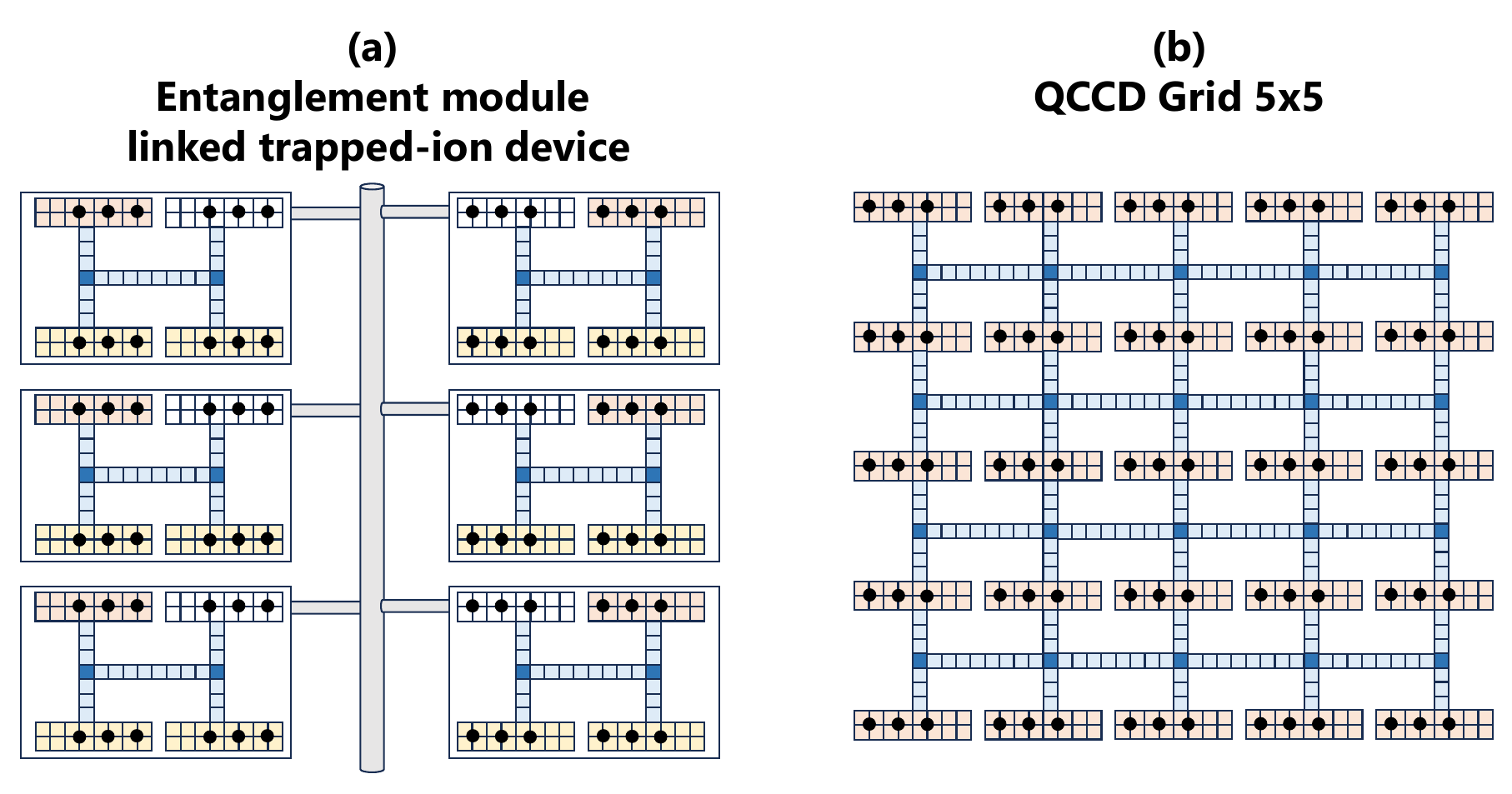}
    \caption{
    Comparison of (a) Entanglement module linked and (b) QCCD grid architectures.
    Entanglement module linked trapped-ion device consists of multiple improved QCCD devices, which are connected by a fiber. Qubits in different traps can interact through the fiber to perform two-qubit gates.
    }
    \label{fig:device compare}
\end{figure}
While the QCCD architecture addresses near-term scalability challenges, it requires integrating optical waveguides into every trapping zone to enable individual qubit control, which presents significant difficulties in chip fabrication and increases the complexity of optical control~\cite{brown2021materials, mordini2025multizone}. Moreover, most quantum algorithms naturally consist of a mix of local and nonlocal gate operations, making the conventional QCCD approach of embedding individual laser systems in every trapping zone inefficient, especially since local operations are only required in specific regions. For non-local operations, the traditional approach of physically shuttling ions over long distances incurs significant latency and exposes qubits to increased motional heating and decoherence. To fully exploit the strengths of the QCCD model and the intrinsic suitability of trapped ions for photonic remote entanglement, an entanglement module linked QCCD (EML-QCCD) architecture (Fig.\ref{fig:device compare}) can be employed as a more scalable and modular alternative.

Trapped ion qubits have narrow linewidth transitions at optical frequencies, making them ideal for low-loss transmission through optical fibers.
This characteristic makes trapped-ion platforms attractive for distributed quantum computing and quantum networking~\cite{o2024fast}. 
Photon-mediated remote entanglement generation and ion-photon interfaces enable the realization of non-local quantum gate operations, which are essential for modular architectures. 
Such entanglement-linked approaches extend the spatial scalability of ion-trap systems and provide a practical pathway toward quantum internet infrastructure and interoperability between heterogeneous quantum platforms~\cite{drmota2023robust, krutyanskiy2023entanglement}.
Recent studies have explored the use of modular quantum architectures, beyond physically shuttling qubits on a microchip~\cite{akhtar2023high}, an increasing number of experiments are now exploring photonic approaches for long-range coupling of distant ions or atoms~\cite{o2024fast, knollmann2024integrated,mordini2025multizone}. 
Additionally, researchers aim to miniaturize and integrate optical waveguides on-chip to deliver photons within trapped-ion architectures~\cite{romaszko2020engineering}.
This approach, akin to the concept of distributed classical computing, allows for the construction of large-scale quantum computers by connecting smaller, independently controlled ion trap units via optical fibers. Such modular systems enable entanglement between different ion registers, providing a viable pathway to scaling quantum hardware beyond current limitations. 

The compilation process in EML-QCCD architectures involves transforming high-level quantum algorithms into executable gate sequences. This requires the development of compilation techniques that not only optimize gate scheduling and placement but also mitigate physical limitations~\cite{saki2022muzzle, kreppel2023quantum, wu2021tilt, wu2024boss, murali2019noise}, as seen in prior work on QCCD devices. However, these existing approaches do not account for the distinct operational zones within QCCD architectures, nor do they consider the effects of fiber based entanglement links. Addressing these additional complexities calls for a specialized compiler tailored to the unique constraints of EML-QCCD systems.


In this paper, we present a specialized compiler designed to optimize shuttle scheduling in EML-QCCD. Our main contributions are summarized as follows:

\begin{itemize}[topsep=10pt]
    \item We develop a \underline{mu}lti-level \underline{s}huttle \underline{s}cheduling framework for large-scale entanglement module linked \underline{t}rapped-\underline{i}on devices, referred to as MUSS-TI. Analogous to multi-level cache management, MUSS-TI takes into account the interactions and effects among different operational zones within the QCCD architecture. 
    \item At the technical level, we design a dedicated SWAP insertion algorithm tailored for EML-QCCD. By strategically placing SWAP gates between different QCCDs, MUSS-TI enables logical qubit transfer across traps, effectively reducing redundant shuttling operations that commonly occur in conventional QCCD architectures.
    \item At the results level, our evaluation shows that MUSS-TI reduces shuttle operations by 41.74\% and shortens execution time by 58.9\% compared to existing methods for small-scale applications. Additionally, MUSS-TI shown to be more suitable for medium- and large-scale applications, achieving reductions in shuttle operations of 73.38\% and 59.82\%.
    \item MUSS-TI also supports EML-QCCD configurations with varying numbers of optical zones and recommends better EML-QCCD trap capacities to achieve better performance across different applications.
\end{itemize}


\section{Background}

\begin{figure*}[t]
    \centering
    \includegraphics[width=0.9\linewidth]{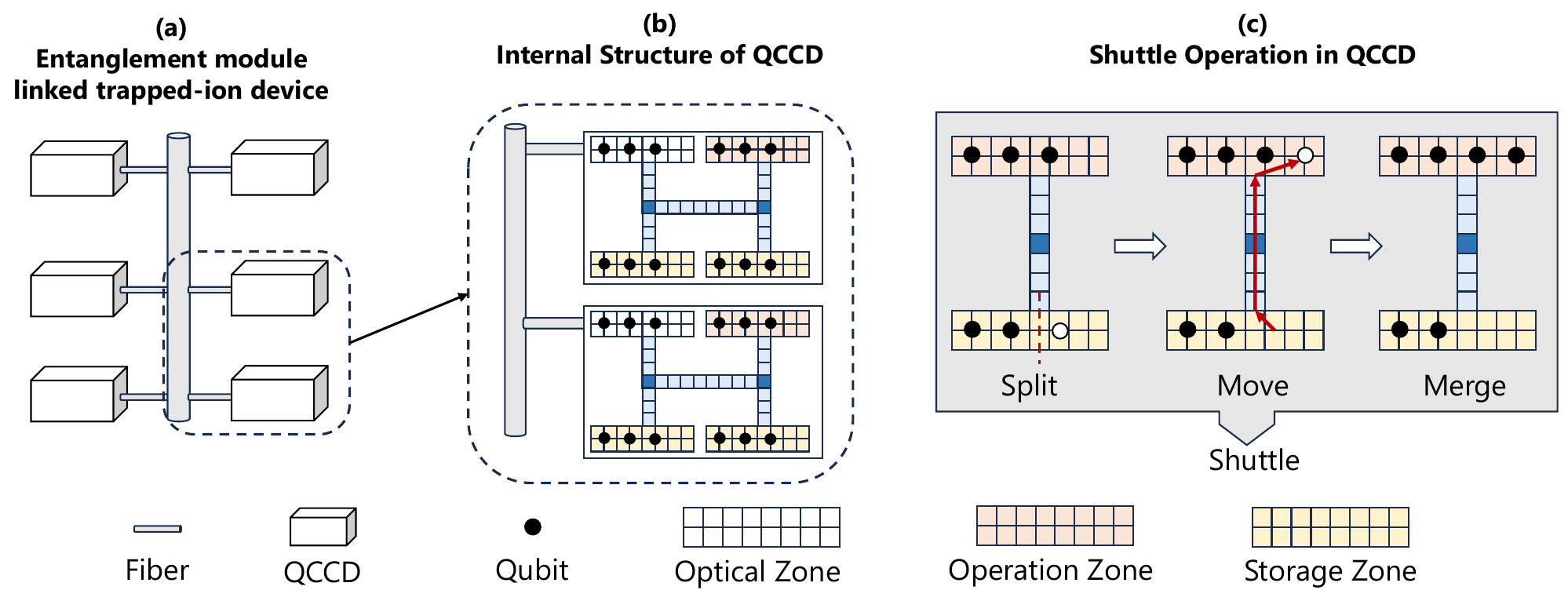}
    \caption{(a) The overview of entanglement module linked QCCD (EML-QCCD) devices. EML-QCCDs are interconnected via fibers, which enable communication between the qubits within different QCCDs.
    (b) The internal structure of EML-QCCD. Yellow traps are storage zones for qubits. Red traps form the operation zone, where qubits are fully connected and can undergo two-qubit gates. White traps represent the optical zone, connected via fibers to enable entanglement and two-qubit gate operations with qubits in other QCCDs' optical zones. Additionally, qubits within the optical zone can also perform two-qubit gates internally, similar to the operation zone.
    (c) Shuttle operation in QCCD. In the QCCD device, to meet the hardware limitations, shuttle operations are required to move qubits between traps. A complete shuttle operation consists of three steps: split, move, and merge.
    }
    \label{fig:ion trap 3.0 devices}
\end{figure*}

\subsection{Introduction to Quantum Computation}
Qubit is the basic unit of quantum information in quantum computing, which has two computational orthogonal basis states $\ket{0}$ and $\ket{1}$.
For a single-qubit state, it can be described by a linear combination of $\ket{0}$ and $\ket{1}$, that is $\ket{\psi} = \alpha\ket{0} + \beta\ket{1}$, where $\alpha$ and $\beta$ are complex amplitudes satisfying $|\alpha|^2 + |\beta|^2 = 1$. 
The $n$ qubit state can be expressed as $\ket{\psi} = \sum_i^{2^n-1} \alpha_i \ket{i}$, where $\alpha_i$ are complex amplitudes satisfying $\sum_i^{2^n-1} |\alpha_i|^2 = 1$. Quantum gates are used to transform quantum states. In trapped-ion systems, the commonly used gates are the M{\o}lmer–S{\o}rensen (MS) gates~\cite{sorensen1999quantum}.

\subsection{Trapped-Ion Preliminary}
\textbf{Trapped-ion Quantum Computing. }
Trapped-ion quantum computing employs individual ions as qubits to execute quantum operations, primarily through single-qubit rotations and entangling MS gates~\cite{haffner2008quantum}. The approach leverages the intrinsic advantages of trapped ions, such as long coherence times and highly stable internal energy levels, which support precise and high-fidelity gate implementations~\cite{harty2014high, gaebler2016high}. 
The electronic structure of the ions enables efficient Doppler cooling in the resolved sideband to the motional ground state, facilitating accurate initialization~\cite{harty2014high}, while fluorescence-based state-dependent detection provides high-fidelity readout~\cite{myerson2008high}. 
The fully connected nature enhances programmability and flexibility for quantum algorithms and QEC.

One of the distinctive strengths of trapped-ion systems lies in the versatility of ion control modalities, including static and RF electric field confinement, fast ion transport via dynamic potential modulation, and laser-driven or microwave-based qubit operations. These diverse methods enable flexible manipulation and modular scaling strategies which we call the QCCD~\cite{wineland1998experimental,kielpinski2002architecture} and have emerged as a pivotal approach, witnessing rapid development in recent years~\cite{wan2019quantum, kaufmann2017scalable, pino2021demonstration}.
The QCCD demonstrates excellent scalability and is considered a key technology for realizing large-scale trapped-ion devices. Recently, a distributed large-scale trapped-ion computing technology named TITAN has been proposed \cite{chu2024titan}, showcasing the powerful potential of QCCD.

\textbf{Entanglement Module Linked Trapped-Ion.}
Traditional QCCD architectures face scalability challenges due to the need for dense on-chip optical waveguide integration and complex laser power routing, which increase fabrication difficulty, thermal noise, and system instability. Embedding laser control at every trap zone is resource-inefficient, especially since local gates are only required in specific regions. 
Moreover, long-range ion shuttling for non-local operations incurs latency and error accumulation. In contrast, photonic interconnects enable more efficient and reliable implementation of non-local gates and remote entanglement, reducing transport overhead and improving modular scalability.

Here, we consider this more scalable trapped-ion architecture, as illustrated in Fig. \ref{fig:ion trap 3.0 devices}.
The improved QCCD devices presented here differ from standard QCCD devices by assigning distinct functions to different traps. 
The yellow traps serve as storage for qubits, while the red traps are operation zones for gate operations. The white traps function as optical zones, enabling two-qubit gate operations with qubits from other QCCD optical zones.
Also, in such architecture, only the red and white trap areas need integrated optical waveguides, compared to the traditional QCCD structure, this significantly reduces the number of optical fibers required for integration, while remote entanglement effectively enhances the swapping efficiency of nonlocal quantum gates.

In summary, this structure offers several key advantages:
\begin{itemize}[topsep=5pt]
    \item Firstly, the optical fiber connections allow us to perform two-qubit gates on two different QCCD devices, distinguishing it from traditional QCCDs and reducing the number of shuttle operations. A similar design is also seen in TITAN \cite{chu2024titan}, but ours has some distinctive features. Moreover, this design is more readily achievable in the near term~\cite{mehta2016integrated,kwon2024multi}.
    \item Secondly, we implement a more refined design of the traps within the QCCD, categorizing them into different ``zones'' such as storage zone, internal operation zone, and fiber entanglement zone. This design not only simplifies the hardware implementation but also circumvents the complex scheduling issues associated with an overly intricate QCCD grid.
    \item Thirdly, such a device inspires us to develop a novel algorithm, which incorporates the multi-level cache scheduling concept from classical computer systems. This development greatly enhances gate scheduling for trapped-ion devices, leading to a substantial reduction in operational overhead.
    \item Additionally, this approach avoids the need to make the QCCD grid excessively large, which would otherwise lead to worse performance. For example, a larger trap capacity can result in lower gate fidelity~\cite{ratcliffe2018scaling}. Moreover, an overly large grid may introduce unnecessary shuttle operations.
\end{itemize}

Compared to the TITAN, EML-QCCD is more achievable in the near future. 
The TITAN approach requires each module to feature dozens of optical ports interconnected through large-scale, ultra-high-speed photonic switches to achieve parallel inter-module entanglement. 
However, this significantly increases cost and complexity: the optical-matter interface is expensive, and scaling the number of ports greatly complicates manufacturing and alignment.
Additionally, current high-speed switches cannot reliably maintain photon entanglement fidelity under millisecond-level switching speeds, creating a practical implementation bottleneck.

EML-QCCD addresses these issues by significantly reducing the number of ports and introducing functional zone partitioning within each module. Specifically, each module maintains only the minimal number of optical ports necessary to meet near-term NISQ-stage entanglement needs. We further divide each module internally into entanglement, operation, and storage zones. 
High-precision optical interfaces are concentrated in the entanglement and operation zones, while the storage zone benefits from mature ion-trap microfabrication techniques, simplifying manufacturing. Cross-module entanglement requires fewer ports, eliminating the need for complex high-speed photonic switches. 

\subsection{Motivation}

\paragraph{Hardware Limitations of Traditional QCCD}
In trapped-ion quantum computing, the optical zone refers to a specific region on the trapped-ion device where light interacts with ions~\cite{mehta2016integrated,kwon2024multi}. 
In this region, lasers are precisely controlled to interact with the trapped ions, allowing for quantum state manipulation and measurement. 
In designing trapped-ion quantum devices, it's essential to consider both the confinement of more ions using electrodes and the evaluation of required laser resources. Efficient access to laser channels that can drive ion internal states is crucial.
Under current technological conditions, qubit manipulation within a QCCD is limited to a specific region. This constraint arises from the finite number of available laser channels and the challenges of arranging them without interference. Therefore, we focus on the EML-QCCD device, reflecting the current state of the hardware.

\textit{Compiler Limitations in Shuttle Scheduling.}
Current QCCD compilers exhibit significant limitations, primarily due to their inefficiency in minimizing shuttle operations in EML-QCCD devices. Traditional QCCD compilers allow two-qubit gates to be applied in arbitrary zones, without accounting for the distinct operational zones in EML-QCCD devices. This limitation and the overlooked properties of EML-QCCD can lead to increased thermal output, which compromises circuit fidelity.

\textit{Fiber entanglement for robust distributed quantum computing networks.}
Fiber entanglement is a technique that aims to connect multiple quantum hardware devices using optical fibers, enabling quantum entanglement across different devices~\cite{mirhosseini2020superconducting,malinowski2023wire}. This technology enables efficient quantum information transfer between different trapped-ion systems through the exchange of flying photons, significantly expanding the scale and connectivity of quantum computing systems.
By generating entangled states via optical fibers, coordinated operation between trapped-ion devices becomes feasible, while modulating one device will not affect others due to the ability to switch fiber entanglement. 

\section{Pipeline for MUSS-TI }
To achieve improved performance of quantum circuits within entanglement module linked trapped-ion architecture, we propose a novel quantum circuit compilation method. We integrates concepts from memory scheduling algorithms to design a strategy that efficiently manages resources and operations.
The architecture of entanglement module linked trapped-ion parallels that of memory scheduling: qubits are analogous to tasks, the storage zone functions as external storage (level 0), the operation zone as internal memory (level 1), and the optical zone serves as the CPU (level 2).
Two-qubit gates require ions to arrive at the correct zones on schedule. When partners are misplaced, our algorithm moves them. If the zones are occupied, resident qubits are evicted—much like a page fault—borrowing classical memory-scheduling techniques for the ion trap.


\subsection{Dependency Graph}
A quantum circuit consists of a sequentially ordered set of quantum operations, which can be represented as a directed acyclic graph (DAG).
The first step of MUSS-TI translates the circuits into a representation that captures the dependencies between gates, commonly known as a dependency graph.
In this representation, each gate is a node, and a directed edge \((g_i, g_j)\) indicates that gate \(g_j\) can only execute after \(g_i\). Nodes with zero in-degree are ready for immediate execution if they satisfy the topological constraints, and they are then removed from the graph. Single-qubit gates are often disregarded in practical implementations, and this simplification will be applied in the following steps.
Constructing the dependency graph for a given sequence of gates is relatively efficient, with a time complexity of \( O(g) \), where \( g \) denotes the total number of gates.

\subsection{Gate scheduling and qubit routing}
\label{sec:compiler framework}

\begin{figure}[h]
    \centering
    \includegraphics[width=1.0\linewidth]{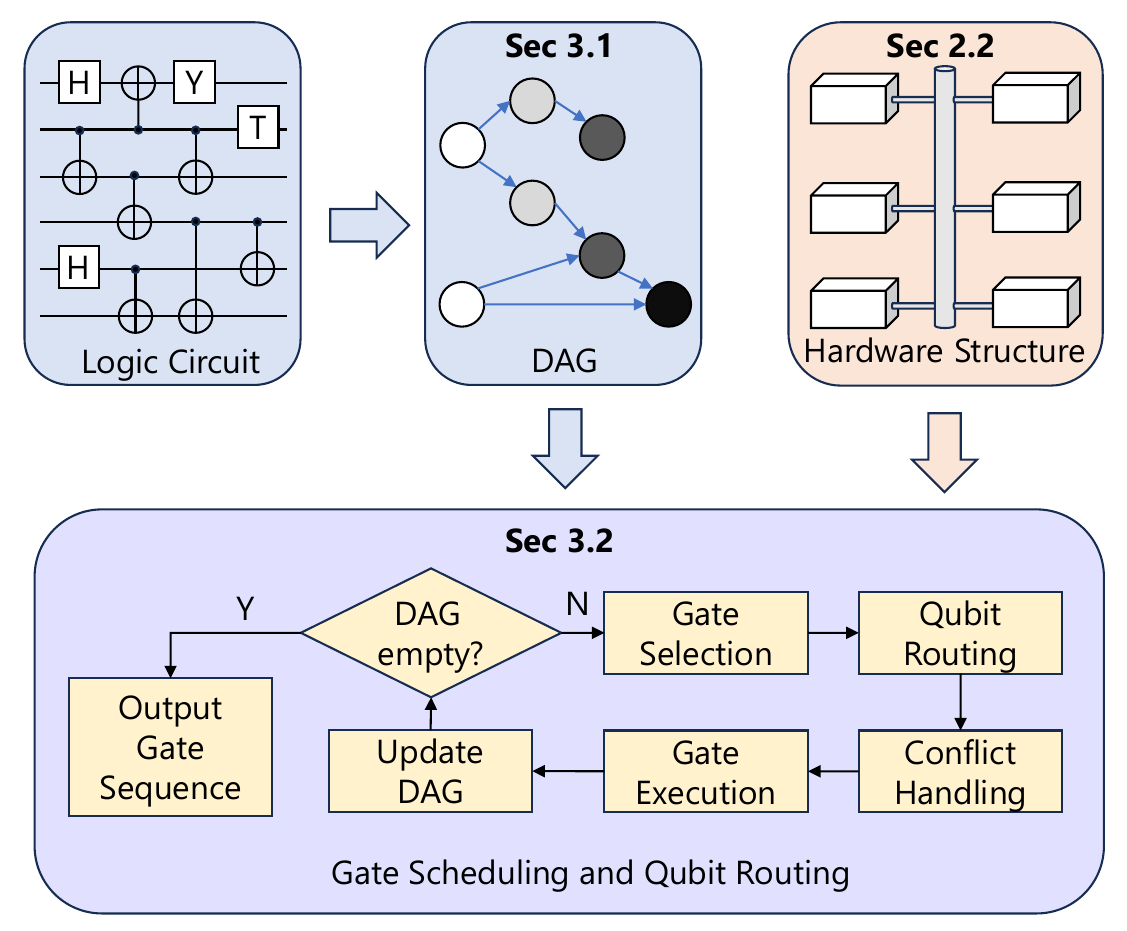}
    \caption{Illustration of the MUSS-TI framework.}
    \label{fig:compiler framework}
\end{figure}

\begin{figure*}[t]
    \centering
    \includegraphics[width=0.9\linewidth]{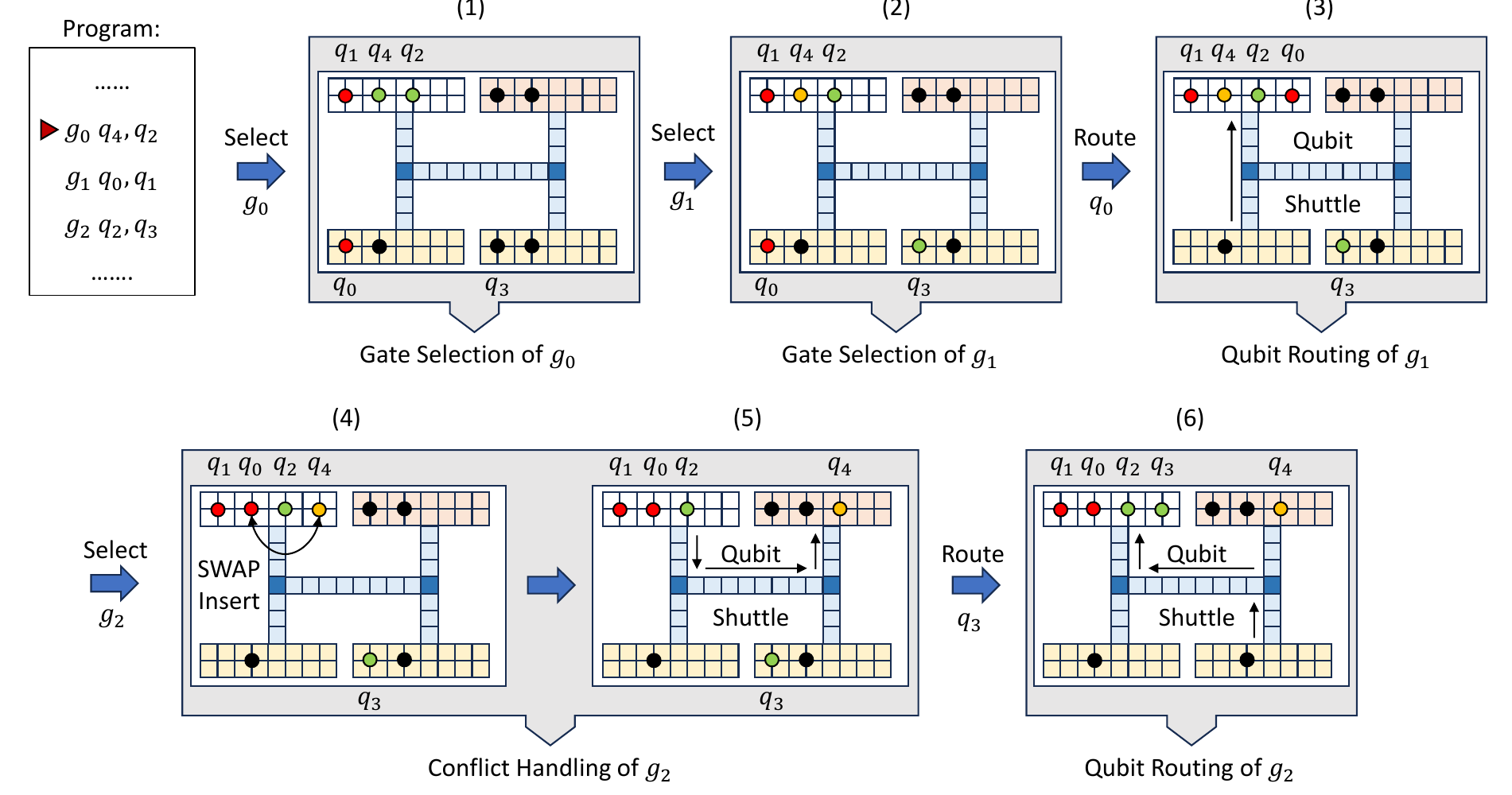}
    \caption{An example of MUSS-TI scheduling. 
    Initially, we execute $g_0$ since it already meets the hardware requirement.
    Given that neither \( g_1 \) nor \( g_2 \) meets the requirements, we select \( g_1 \) on a first-come, first-served basis. Subsequently, we aim to pick a zone for \( g_1 \).
    Both zones at level 1 and level 2 are viable options; however, due to the shorter distance, $q_1$ is relocated to the level 2 zone where $q_0$ resides.    
    Similarly, we select level 2 zone for $g_2$ and $q_3$ needs to be moved to the level 2 zone occupied by $q_2$. 
    At this point, the level 2 zone has reached its capacity limit. Given that $q_0$ and $q_1$ have been recently utilized, we apply the LRU strategy to evict $q_4$ from the level 2 zone. In accordance with the Multi-level scheduling strategy, $q_4$ is then transferred to the level 1 zone.
    It is important to note that shuttling operations can only occur at the edges of the qubit chain. Therefore, to facilitate this movement, we employ SWAP gates to reconfigure the arrangement of the qubit chain.
    }
    \label{fig:example of cache like scheduling}
\end{figure*}
Inspired by memory scheduling algorithms, we propose a compiler tailored for EML-QCCD. We substantiate this claim in the subsequent experimental section, where we demonstrate the superiority of our method. The overall algorithmic workflow is illustrated in Fig.~\ref{fig:compiler framework}.
The key components of our algorithm are gate selection, qubit routing, and conflict handling, each plays an important role in improving the efficient execution of quantum circuits.

\textbf{Gate Selection:} This process focuses on selecting a gate from the frontier of the DAG, where multiple gates may be available. Establishing a good execution order is essential to minimize overhead and enhance execution efficiency. By prioritizing effective gate selection, potential conflicts can be mitigated, thereby improving the performance of quantum circuit execution.

\textbf{Qubit Routing:} This involves scheduling the qubits required by the selected gate and moving them to the proper position for execution. The goal is to determine the best zone for qubit placement while minimizing associated costs.

\textbf{Conflict Handling:} This process manages situations during qubit routing where the selected zone is already occupied. Effective conflict resolution strategies are essential to address these issues and ensure the correct execution of quantum circuits.

To address these components, we introduce the MUSS-TI framework. An example is illustrated in Figure \ref{fig:example of cache like scheduling}. 
Our algorithm employs the following key strategies to optimize the scheduling process:

\textbf{Prioritize executable gates.} In the frontier layer of a DAG, multiple two-qubit gates may be present, though not all are immediately executable. We prioritize executing those that can be executed right away to minimize unnecessary shuttle operations and enhance overall efficiency. If no such gates are available, we schedule based on a first-come, first-served principle. The computational complexity of this step is \( O(n) \), where \( n \) is the number of qubits in the circuit.

\textbf{Multi-level scheduling.} When routing qubits associated with the selected gate to a zone that meets execution requirements, it is crucial to choose an appropriate target zone for transfer. Our algorithm prioritizes zones that are both available and closest in level, ensuring efficient resource utilization and good system performance. This strategy is also employed for conflict handling, where we use it to select suitable zones for scheduling qubits that need to be relocated. The computational complexity of this step is \( O(z) \), where \( z \) is the number of zones in the single device.

\textbf{Qubit replacement scheduler.} This algorithm replaces the qubit that has been inactive the longest, based on the principle of locality, which suggests that this qubit is the least likely to be used soon. In our algorithm, we explicitly employ an LRU (Least Recently Used) based replacement policy, inspired by memory scheduling strategies, to effectively manage qubit replacement. This approach not only enhances scheduling efficiency but also demonstrates near-optimal performance by prioritizing eviction of qubits that have remained unused for the longest duration. The computational complexity of this step is \( O(c) \), where \( c \) is the zone capacity. 
Therefore, the overall time complexity of the algorithm is $O(g\cdot(n+z+c))=O(gn)$.

\subsection{SWAP Gate Insertion}
\label{swap gate insertion}

\begin{figure*}[t]
    \centering
    \includegraphics[width=\linewidth]{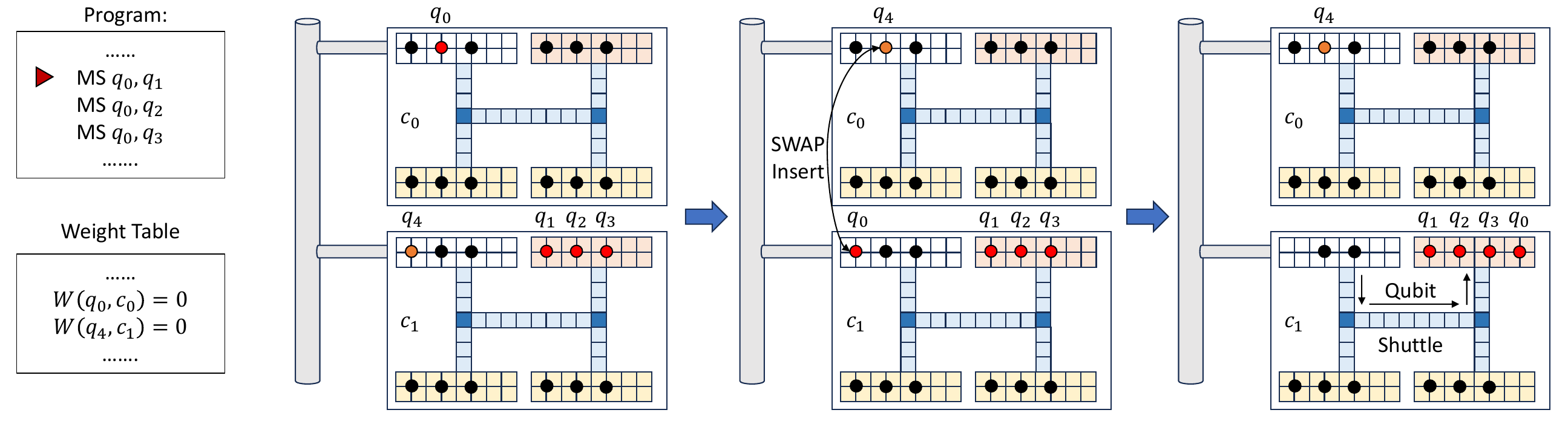}
    \caption{An example of SWAP gate insertion. By inserting a SWAP gate to transfer qubit $q_0$ to another trap, the number of shuttle operations can be effectively reduced.
    Moreover, the operational cost of two-qubit gates in the internal zone is significantly lower than that of fiber-entanglement operations in the optical zone. 
    }
    \label{fig:ion swap insertation}
\end{figure*}

In the Entanglement Module Linked Trapped-Ion architecture, we only consider inserting SWAP gates between qubits located on different QCCDs. For instance, as illustrated in Fig. \ref{fig:ion swap insertation}, performing a swap operation between two qubits (logically) can reduce the number of subsequent shuttles required. Consider a scenario where two-qubit gates need to be applied between $q_0$ and each of $\{q_1,q_2,q_3\}$. 
Using conventional scheduling methods would require at least three shuttles to move the necessary qubits into the optical zone. However, by strategically inserting a SWAP gate, this number can be reduced to just one shuttle.
Furthermore, it is generally observed that the fidelity of two-qubit gates executed within the operation zone tends to be superior to that of fiber-entangled gates. This inherent advantage can significantly enhance the overall fidelity performance.

Our SWAP insertion approach differs from existing methods in two key ways. First, it is not just a scheduling task to meet hardware constraints. Second, the inserted SWAP gates significantly enhance qubit connectivity. Thus, a hardware-specific SWAP insertion method is needed.
To address it, we introduce a weight table \( W \) to assess the necessity and suitable placement of SWAP gates. Here, \( W(q_i, c_j) \) represents the number of gates involving qubit \( q_i \) and qubits located on QCCD \( c_j \) within the first \( k \) layers of the DAG. We set \( k \) to 8, allowing for a look-ahead capability.
After executing a two-qubit gate involving qubits  $q_a, q_b$ on different QCCDs, we check if $W(q_a, c_a)$ is zero. If true and $W(q_a, c_j) >T$ for another QCCD $c_j$ with a qubit $q_c$ such that $W(q_c, c_j)=0$, we insert a SWAP gate to swap these two qubits. 
Because a SWAP gate is typically composed of three MS gates, $T$ should not be less than 3.
In our later experiments, $T$ is a predetermined threshold that we set to 4.
As for $q_b$, we perform a similar operation.

Given that qubits may not be utilized in every layer of the DAG, \( k \) should not be excessively small. 
If \( k \) is excessively large, SWAP insertion may be canceled due to the qubit still being required in the current trap in the near future. Therefore, we selected \( k = 8 \).
The time complexity of determining whether to insert a SWAP gate is \( O(kn) \), where \( n \) denotes the number of qubits.
The selection of the forward-looking ability \( k \) is independent of the application size. Alternatively, with an understanding of the locality of the input circuits, adjustments to \( k \) can be made.

\subsection{Initial Mapping}
\label{pre mapping}
\textbf{Trivial Mapping.} It is widely acknowledged that the initial mapping has a significant impact on the overall performance of a quantum circuit.
Given that zones with higher levels typically offer superior functionality, it is reasonable to prioritize the placement of qubits in zones ordered by their levels from highest to lowest. This sequential placement of qubits, based on their order, constitutes the approach that we designate as the trivial mapping.

\textbf{SABRE.} We also develop a two-fold search method analogous to previous work~\cite{li2019tackling} to achieve a better qubit placement. 
We first transform the input circuit into a DAG representation \(\mathcal{G}\), and then initialize the mapping with a trivial mapping \(\pi\). Subsequently, we execute \(\mathcal{G}\) based on this initial mapping. After executing \(\mathcal{G}\), we obtain a final mapping \(\pi'\). We then use \(\pi'\) as the initial mapping to execute the reversed graph \(\mathcal{G'}\), where ``reverse'' refers to inverting the direction of all edges in \(\mathcal{G}\). We use the final mapping \(\pi''\), obtained after executing \(\mathcal{G'}\), as the initial mapping to formally execute \(\mathcal{G}\).
The mappings obtained from these two executions serve as the initial mapping for further optimization.
This process acts as a pre-loading mechanism for memory blocks, proactively loading qubits into the working area before use. 
The core idea was introduced in~\cite{li2019tackling}, and we refer to this mapping as SABRE.
Previous studies on superconducting devices with limited qubit connectivity show that initial mappings from two-fold search strategies often yield minimal differences. In contrast, the enhanced connectivity of EML-QCCD architectures amplifies the effects of two-fold search, resulting in more significant variations in initial mappings.

\section{Experimental Setup}

\textbf{Baseline algorithms.} We primarily compare our algorithm with three benchmark approaches. Specifically, we select the widely recognized QCCD simulator~\cite{murali2020architecting_iontrap} and a recently proposed compilation algorithm~\cite{dai2024advanced} as key baselines. The former is a standard benchmark for trapped-ion simulators, while the latter focuses on reducing the number of shuttling operations. 
Additionally, we include the open-source Munich Quantum Toolkit (MQT)~\cite{schoenberger2024shuttling}, which offers a compilation method for scalable QCCD architectures.

\textbf{Benchmark Applications.} These benchmark applications span from circuits utilizing nearest-neighbor gates and short-distance patterns, which are typically optimized for near-term superconducting devices, to well-known algorithms that illustrate quantum advantages. These benchmark circuits are selected based on the previous work~\cite{murali2020architecting_iontrap} and from QASMBench~\cite{li2022qasmbenchlowlevelqasmbenchmark} with a qubit number ranging from 30 to 299 and a 2-qubit gate number ranging from 31 to 4376. 



\textbf{Simulation Devices. } All experiments were conducted on a Windows 11 system with Intel Core i7 processor (2.4 GHz and 16GB RAM) using Python 3.9. 

\textbf{Fidelity Model. } 
Within a single device, the ion dynamics can be aligned with the near-term QCCD framework, encompassing four operations: Split, Swap, Move, and Merge~\cite{murali2020architecting_iontrap}. For ion shuttling-based quantum operations, two primary factors limit the fidelity of quantum gates. First, shutting qubits introduces additional time overhead, which becomes detrimental under finite qubit lifetimes, reducing the overall gate fidelity. Second, repeated shuttling operations induce motional heating that accumulates linearly with the number of shuttles, further degrading system performance. To quantify the impact of these effects, we estimate the fidelity of shuttle operations using the following formula:
\begin{align}
    F = e^{-t/T_1} -k\Bar{n},
\end{align}
where \( T_1 = 600 \times 10^6 \, \mu \text{s} \)  represents the qubit lifetime \cite{wang2020single}, and \( k = 0.001 \) denotes the ion trap heating rate. The shuttle operation induces heat accumulation in its corresponding zone, which subsequently leads to a decrease in the fidelity of subsequent gate operations. We denote the background fidelity of zone \(i\) as \(B_i\). The fidelity of gate $g$ acting on this zone can be represented as \(F'_g = B_i F_g\).

In trapped-ion systems, qubits are encoded in two internal atomic energy states. The natural availability of atomic energy levels with long-lived excited states, as well as the use of degenerate levels within hyperfine structures, enables significantly prolonged qubit lifetimes. Recent technological advancements in trapped-ion platforms have demonstrated qubit lifetimes exceeding one hour, highlighting the robustness and reliability of this system for quantum computing applications~\cite{wang2021single}. Regarding the heating rate, the QCCD structure requires frequent ion transport across different trap locations, necessitating rapid atom movement. 
This dynamic increases atomic kinetic energy and, consequently, a rise in temperature, which adversely impacts the fidelity of quantum operations. Therefore, this additional term is included in the fidelity estimation formula to account for such effects.

Moreover, due to the quadratic growth in modulation pulse complexity required for multi-qubit control, the two-qubit gate fidelity exhibits a quadratic decay with respect to the number of ions in the trap, and the decay coefficient $\epsilon$ can be treated as a tunable parameter reflecting the precision of experimental instrument~\cite{martinez2022analytical}.
As for gate operations, we set $\epsilon=1/25600$ and $N$ represents the number of qubits currently in the trap.
The relevant parameters for these operations are detailed in Table~\ref{tab:op parameters}.

\begin{table}[h]
    \centering    
    \caption{This table presents the relevant parameters in MUSS-TI, including both time and fidelity~\cite{blakestad2009high,walther2012controlling,shu2014heating,gutierrez2019transversality,wu2018noise}.}
    \begin{tabular}{|c|c|c|c|}    
         \hline
         \textbf{Type} &\textbf{Operation} & \textbf{Time}& \textbf{Fidelity} \\
         \hline
         \multirow{4}{*}{Trap}&Split & 80 $\mu s$& $\Bar{n}=1$ \\
         &Move & 2$\mu m/\mu s$ & $\Bar{n}=0.1$ \\
         &Swap & 40  $\mu s$ & $\Bar{n}=0.3$ \\
         &Merge & 80 $\mu s$ & $\Bar{n}=1$\\
         \hline
         \multirow{3}{*}{Gate}&1-qubit gate & 5 $\mu s$ & 0.9999\\
         & 2-qubit gate & 40 $\mu s$ & 1-$\epsilon N^2$\\
         &Fiber entangle  & 200 $\mu s$ & 0.99\\
         \hline
    \end{tabular}
    \label{tab:op parameters}
\end{table}

\textbf{Metrics.} (1) \textit{Shuttling} refers to the process of moving a qubit from one trap to another. Since it causes heat accumulation in the trap, which affects subsequent gate operations, it is important to minimize the shuttle count to maintain overall circuit fidelity. (2) \textit{Circuit execution time} also impacts overall fidelity and will be taken into account in our evaluation. (3) \textit{Fidelity} (reliability) is affected by both shuttling and execution time and is computed as the product of individual gate fidelities. This approach is widely adopted and has been experimentally validated in prior works~\cite{murali2020architecting_iontrap, wright2019benchmarking, linke2017experimental, murali2019noise}.

\begin{table*}[h]
    \centering
    \caption{Comparison on small-scale applications. The trap capacity of Grid 2$\times$2 is 12, whereas the trap capacity of Grid 2$\times$3 is 8.}
    \begin{tabular}{
    |c|c|c|c
    |c|c|c|c|c
    |c|c|c|c|c|}
        \hline
         \multirow{2}{*}{\textbf{Structure}} & \multirow{2}{*}{\textbf{Application}}  & 
         \multicolumn{4}{c|}{\textbf{Shuttle Count}} & 
         \multicolumn{4}{c|}{\textbf{Execution Time}} & 
         \multicolumn{4}{c|}{\textbf{Fidelity}}  \\
         \cline{3-14}
         & & \cite{murali2020architecting_iontrap} & \cite{dai2024advanced} & \cite{schoenberger2024shuttling} & Ours & \cite{murali2020architecting_iontrap} & \cite{dai2024advanced} & \cite{schoenberger2024shuttling} & Ours & \cite{murali2020architecting_iontrap} & \cite{dai2024advanced} & \cite{schoenberger2024shuttling} & Ours 
         \\
         \hline
         \multirow{5}{*}{Grid 2$\times$2}&Adder\_32  & 73 &55&187 & \textbf{7} 
         &52700 & 38860& 38960&\textbf{11160}
         &3.7e-05 &7.7e-04 & 2.7e-06&\textbf{0.13}\\
         &BV\_32  & 5 &6& 115 &\textbf{4}
         &5735 &4340 & 26520&\textbf{2160 }
         &0.68 &0.73 & 0.29 &\textbf{0.80}\\
         &GHZ\_32  & 3  & 3&  73 &\textbf{2}
         &4785 &3620 &15000 &\textbf{1760 }
         & 0.71 &0.73 &0.52 &\textbf{0.82}\\
         &QAOA\_32  & 23 &19 & 89&\textbf{14}
         & 31390& 13580&23560 &\textbf{4720}
         &7.7e-03 & 0.15& 0.11&\textbf{0.38}\\
         &QFT\_32  & 84 &80&705 & \textbf{60} 
         &83360 &79500 &180880 &\textbf{47760}
         &4.2e-16 &6.5e-15 &6.0e-79 &\textbf{5.9e-13}\\
         &SQRT\_30  & 155 &111&274 & \textbf{95}
         &127195 &64640 & 69240& \textbf{35080}
         &6.3e-20 &3.6e-09 &6.5e-12  & \textbf{7.7e-07}
         \\
         \hline
         \multirow{5}{*}{Grid 2$\times$3}&Adder\_32 & 103 & 82 &294 &\textbf{10}
         &67265 & 45040&60640 & \textbf{13600 }
         & 7.3e-06&2.9e-04 &2.4e-06 & \textbf{0.27}\\
         &BV\_32  & 9 &13 & 171&\textbf{ 6}
         & 7825& 4960&20320 & \textbf{2640 }
         & 0.61&0.73  &0.45  & \textbf{0.89}\\
         &GHZ\_32  & 5 &6 & 104&\textbf{ 4}
         & 5895&5840 &23840 &\textbf{2320 }
         &0.66  & 1.9e-02&0.50  &\textbf{ 0.90 }
         \\
         &QAOA\_32  & 35 & 28&213 & 32
         & 30075&9160 &44520 & \textbf{5280 }
         & 3.7e-03&2.1e-02 &1.9e-02 &\textbf{0.40}
         \\
         &QFT\_32  & 154 &202&1386 &\textbf{123}
         & 85255& 86340&258520 & \textbf{50040 }
         &4.8e-23  & 2.9e-21&2.2e-85 &\textbf{ 5.5e-16}
         \\
         &SQRT\_30  & 167 & 241 & 712&\textbf{117}
         &125585 &86880 &167360 & \textbf{43920}
         &2.5e-16 & 2.3e-20& 3.7e-16& \textbf{2.5e-05}
         \\
         \hline
    \end{tabular}
    \label{tab:shuttle comparison}
\end{table*}

\textbf{Architecture Setting.} For \textit{small-scale applications} (30–32 qubits), we use 2$\times$2 and 2$\times$3 grid configurations of QCCD as test setups to highlight the advantages of our algorithm. 
We apply MUSS-TI on these standard QCCD structures, highlighting its potential for practical application on existing QCCD systems. 
For \textit{medium-scale applications} (117-128 qubits) and \textit{large-scale applications} (256-299 qubits), we use a 3$\times$4 QCCD grid and a QCCD 4$\times$5 grid as our benchmarks for architectural comparison, respectively.
In our simulation of MUSS-TI, the trap capacity is set to 16, which is consistent with the benchmark. 
Each QCCD in MUSS-TI consists of one optical zone, one operation zone, and two storage zones. Additionally, we impose a constraint that limits the maximum number of qubits per QCCD to 32.
In MUSS-TI, the number of QCCD devices is dynamically adjusted according to the application size.
Specifically, a new 2$\times$2 QCCD grid is added only when the total qubit count exceeds a multiple of 32.

\begin{figure*}[t]
    \centering
    \includegraphics[width=\linewidth]{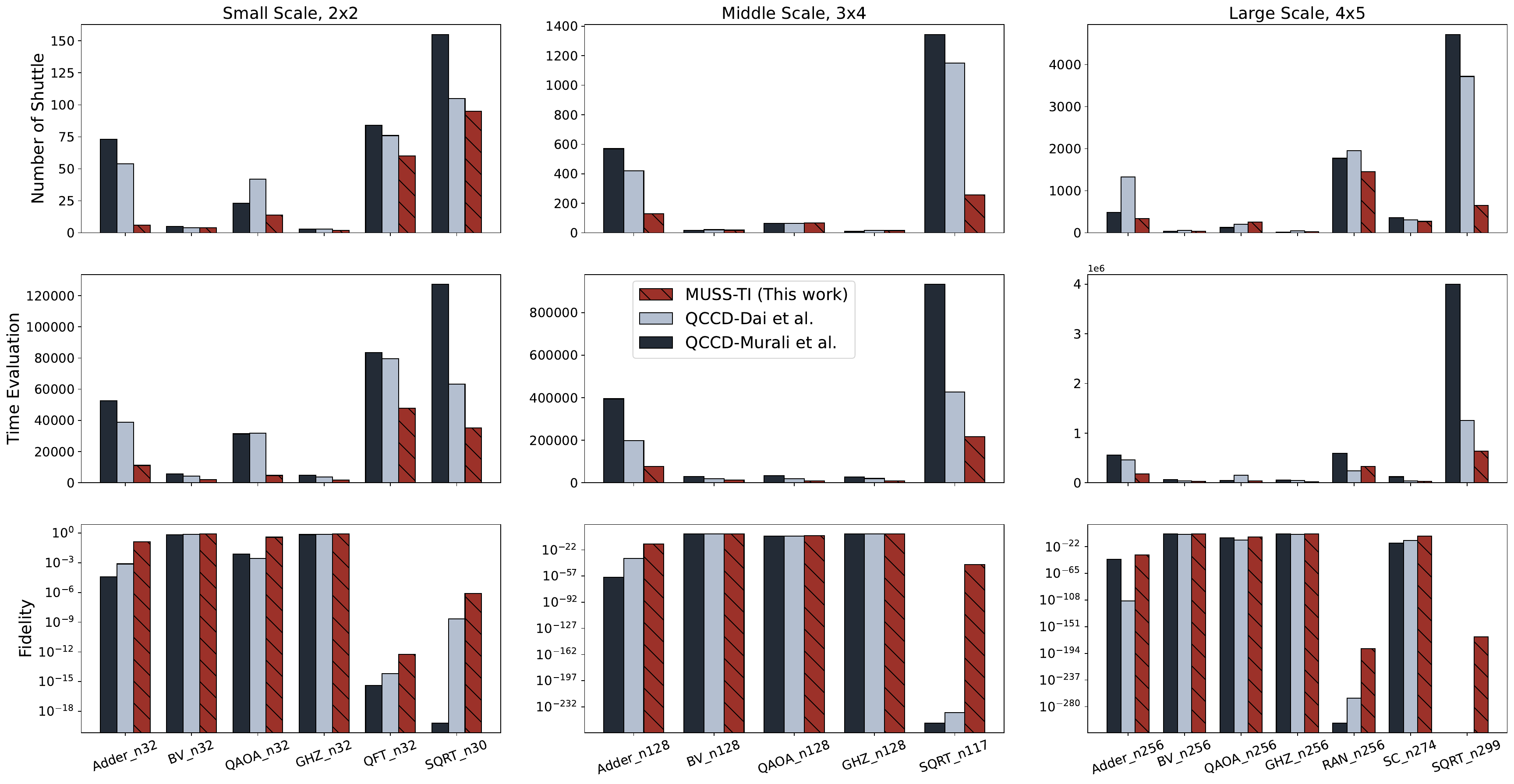}
    \caption{Experimental evaluation of shuttle counts, execution time, and fidelity across various benchmarks provides a comprehensive comparative analysis of our approach against prior studies~\cite{murali2020architecting_iontrap, dai2024advanced}. The results demonstrate significant improvements across multiple large-scale applications.
    }
    \label{fig:ion trap 3.0 results}
\end{figure*}
\section{Experimental Results}

\subsection{Small-scale Application Analysis}
\label{subsec:small scale app analysis}

In Table \ref{tab:shuttle comparison}, we show that our approach significantly reduces shuttle operations compared to previous algorithms for small-scale circuits.
We use three baseline algorithms for benchmarking and select the QCCD grid configurations of 2×2 and 2×3, with trap capacities of 12 and 8, respectively. 
MUSS-TI shows an average improvement of 77.6\% in shuttle operations on the 2$\times$2 grid and 79.45\% on the 2$\times$3 grid.
The reduction in shuttle operations for small-scale circuits is due to the combined use of the LRU strategy and bidirectional mapping. This mapping allows the LRU scheduler to consider both historical and anticipated qubit usage patterns, leading to more efficient scheduling decisions.
Consequently, this reduction leads to shorter execution times and improved application reliability.
The 2$\times$3 grid configuration achieves higher fidelity than the 2$\times$2 grid, despite more shuttle operations and longer execution times.
This outcome mainly results from the smaller trap capacity in the 2$\times$3 grid, leading to higher two-qubit gate fidelities (Table~\ref{tab:op parameters}).
For communication-heavy circuits like \text{QFT}$\_$32, however, the added shuttles can outweigh this gain and reduce fidelity.

\begin{figure*}[t]
    \centering
    \includegraphics[width=1\linewidth]{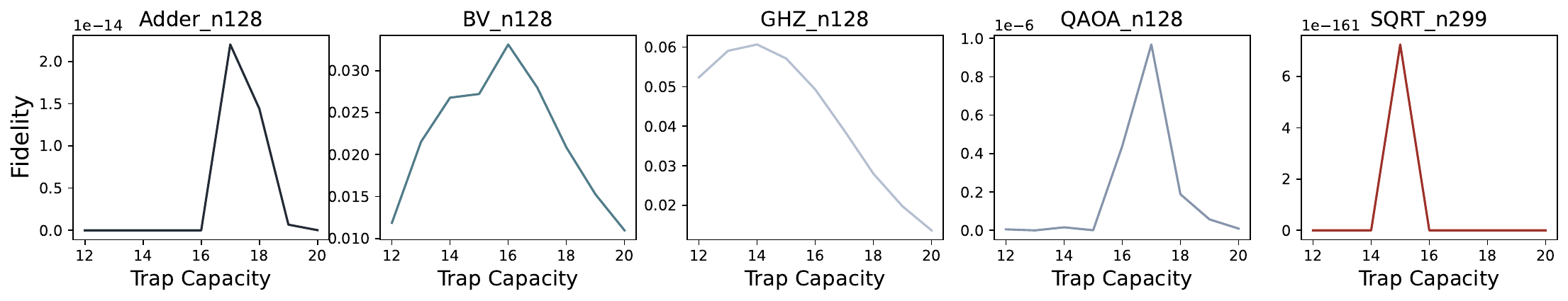}
    \caption{Analysis of EML-QCCD trap capacity. We show that improved EML-QCCD architecture design can enhance the final reliability of quantum applications.}
    \label{fig:trap_capacity}
\end{figure*}

\subsection{Architectural Comparison}

\textit{Number of Shuttle Operations.}
Fig.~\ref{fig:ion trap 3.0 results} (top row) presents a comprehensive comparison of shuttle counts across different scenarios. The results clearly show that MUSS-TI consistently reduces the number of shuttle operations. For small-scale applications, MUSS-TI achieves an average reduction of 41.74\%. The improvement is even more significant for medium- and large-scale applications, with average reductions of 73.38\% and 59.82\%, respectively. The reduction in shuttle operations becomes more pronounced as the application size increases, especially when compared to larger monolithic QCCD devices. This advantage arises from the EML-QCCD architecture's use of a dedicated optical zone, which avoids ion shuttling across multiple QCCDs and enables more efficient communication. For applications with low communication demands, such as QAOA, the benefit is less significant. However, in communication-intensive applications like SQRT, our method achieves a substantial improvement of over 90\%. These results demonstrate the effectiveness of EMI-QCCD not only in small-scale experiments but also in its scalability to large-scale quantum applications.

\textit{Time Evaluation.}
Fig.~\ref{fig:ion trap 3.0 results} (middle row) shows that our method significantly improves execution time estimates across most applications. This improvement is primarily due to the reduction in shuttling operations, as execution time is largely proportional to the number of such operations. For small-scale applications, the average improvement is 58.9\%. The benefits become even more pronounced at larger scales, with MUSS-TI achieving improvements of 64.9\% and 60.3\% for medium- and large-scale applications, respectively. These results highlight that reducing shuttling operations is key to accelerating circuit execution on EML-QCCD devices.

\textit{Fidelity Evaluation.}
Fig.~\ref{fig:ion trap 3.0 results} (bottom row) presents the fidelity estimates for these applications. Note that the fidelity values for QFT are omitted for medium- and large-scale instances, as the circuit contains too many gates. The resulting fidelity values fall below Python’s floating-point precision limit (approximately 2.2$\times 10^{-308}$) and are thus rounded to zero. For applications with simple communication patterns, such as BV, QAOA, and GHZ, the fidelity remains consistent across both devices. However, for more complex applications where the circuit depth and gate count increase, MUSS-TI demonstrates a significant improvement in fidelity compared to existing near term QCCD architectures. 
These gains stem from the reduced number of shuttle operations in MUSS-TI, resulting in higher overall fidelity. This advantage becomes more pronounced as circuit complexity increases, emphasizing the benefits of executing large-scale applications on EML-QCCD.

\begin{figure*}[t]
    \centering
    \includegraphics[width=1\linewidth]{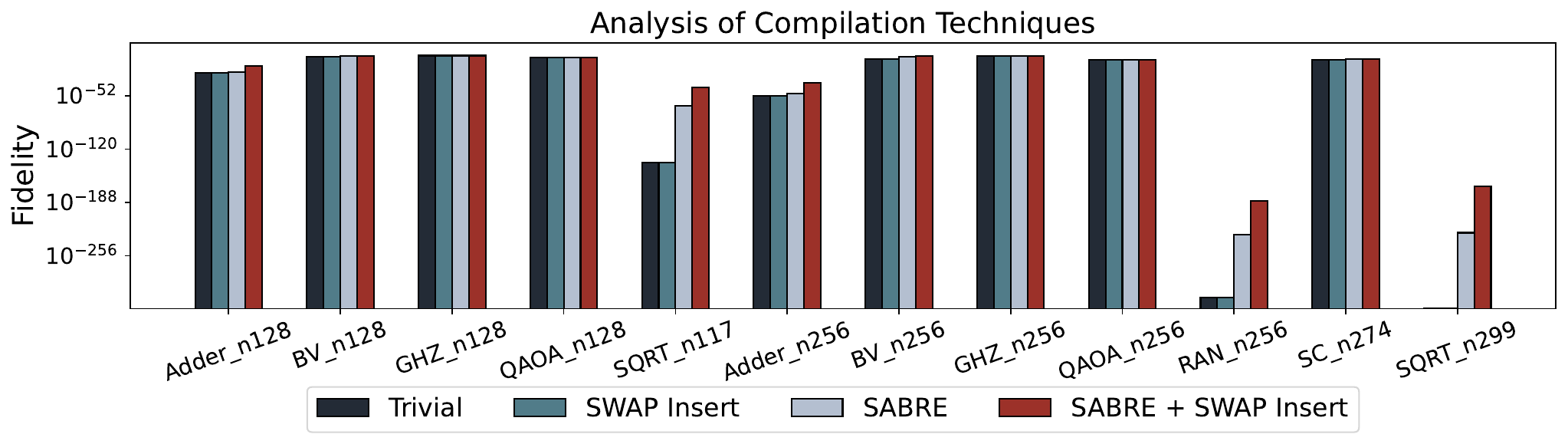}
    \caption{Ablation analysis of the compilation techniques in our proposed compiler MUSS-TI. 
    Trivial refers to the baseline approach without any additional techniques. SWAP insert uses the gate insertion technique described in Section~\ref{swap gate insertion}.  SABRE applies the mapping strategy introduced in Section~\ref{pre mapping}. SABRE+SWAP insert combines both the mapping and gate insertion techniques.
    }
    \label{fig:technique_comparison}
\end{figure*}

\subsection{Trap Capacity Analysis}

In near-term QCCD device design, trap capacity significantly affects the reliability of quantum applications~\cite{murali2020architecting_iontrap}. We also investigate this phenomenon in the context of EML-QCCD. The results (in Fig.~\ref{fig:trap_capacity}) are based on four medium-scale applications with qubit sizes ranging from 117 to 128. 
A common trend in these applications is a fidelity peak at an optimal trap capacity, while both smaller and larger capacities tend to degrade fidelity.
Small trap capacities lead to increased shuttling, which heats the trap and reduces gate fidelity. On the other hand, large capacities reduce shuttle operations but degrade two qubit gate fidelity, as the increasing number of ions makes it more difficult to decouple the phonon modes that mediate qubit interactions.
Overall, trap capacity should be co-designed with the target application. A general range of 14 to 18 qubits per trap tends to yield consistently better fidelity in EML-QCCD, which is slightly below the range of 15 to 25 ions suggested for near-term QCCD devices~\cite{murali2020architecting_iontrap}.

\subsection{Analysis of Compilation Techniques}

To assess the impact of individual compilation techniques within the MUSS-TI framework, we perform an ablation analysis across four configurations. The experimental results are shown in Fig. \ref{fig:technique_comparison}.
The \texttt{trivial} configuration serves as the baseline, utilizing only the basic mapping and the scheduling algorithm described in Section~\ref{sec:compiler framework}, without incorporating any additional optimization techniques. When applying the \texttt{SWAP Insert} technique (described in Section~\ref{swap gate insertion}), we observed only limited improvements in fidelity. 
This technique operates under strict constraints, permitting SWAP insertion only for qubit no longer needed on the current QCCD but required for future operations on another QCCD. In trivial mapping, which serves as our baseline, the number of such qubit pairs is inherently small. Consequently, applying SWAP Insert in isolation results in only marginal fidelity gains.

In contrast, applying the \texttt{SABRE} mapping strategy (introduced in Section~\ref{pre mapping}) produces a more effective initial qubit placement, resulting in an improvement in fidelity. This underscores the critical role of optimized qubit mapping in enhancing overall circuit performance. The combined approach \texttt{SABRE+SWAP Insert}, which integrates both the mapping strategy and the gate insertion technique, yields the most significant improvement in fidelity. 

\begin{figure*}[t]
    \centering
    \includegraphics[width=\linewidth]{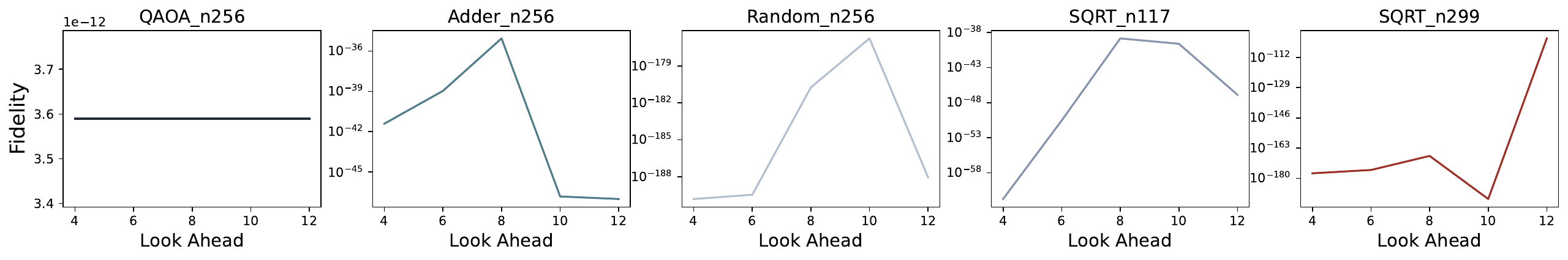}
    \caption{Analysis of look-ahead ability.
    The experimental results demonstrate that the optimal look-ahead ability $k$ varies across different applications, which is attributed to the inherent characteristics of each application.
    }
    \label{fig:look_ahead_analysis}
\end{figure*}

\subsection{Look-ahead analysis.}
\label{sec:look_ahead_analysis}

In the following analysis, we investigate the impact of look-ahead ability on performance. The experimental results are shown in Fig. \ref{fig:look_ahead_analysis}.
Due to the fact that shuttling operations can degrade the fidelity of subsequent quantum circuits, the insertion of SWAP gates in MUSS-TI has a more pronounced impact on circuit performance compared to other existing methods. 
Given the difficulty in precisely estimating the specific impact of inserted SWAP gates on fidelity, we adopt a conservative insertion strategy. Specifically, we only insert SWAP gates between two qubits that are no longer required by the current QCCD in the near future. 
This implies that the look-ahead ability \( k \) is not necessarily better when larger. An excessively large \( k \) may lead to the discovery that a qubit is still needed, thereby forgoing the insertion of a SWAP gate and potentially worsening the overall circuit performance.
The experimental results in Fig. \ref{fig:look_ahead_analysis} further corroborate this observation. Overall, applications with longer qubit communication distances tend to benefit from larger $k$. 
In contrast, nearest-neighbor applications like QAOA are essentially unaffected by changes in \( k \).

\subsection{Scalability Analysis of Compilation Time}

\begin{figure}[h]
    \centering
    \includegraphics[width=0.8\linewidth]{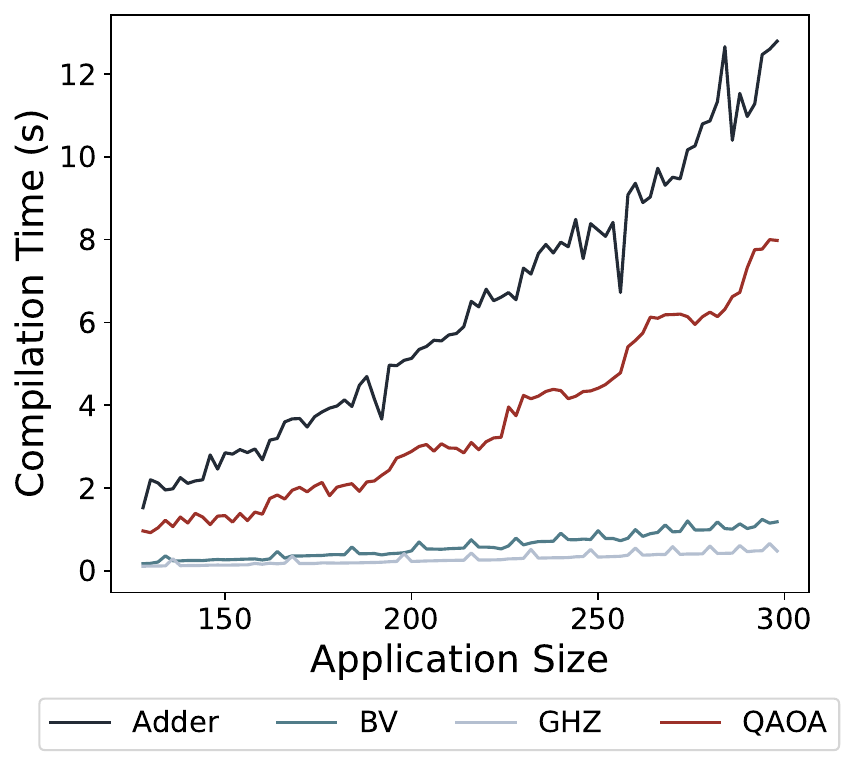}
    \caption{Compilation time analysis of MUSS-TI. }
    \label{fig:compilation_time}
\end{figure}

We then analyze the scalability of compilation time, which is an important part of the classical workload for next generation quantum hardware~\cite{nation2025benchmarkingperformancequantumcomputing}. Specifically, we report the compilation time of MUSS-TI for large scale applications, including Adder, BV, GHZ, and QAOA. The results are shown in Fig.~\ref{fig:compilation_time}.

The time complexity of our algorithm is $O(ng)$, where $n$ denotes the number of qubits and $g$ the number of gates, indicating linear scaling in the worst-case handling of conflict gates. Although compilation time generally increases with application size, it does not exhibit exponential growth. Moreover, due to differences in circuit structure, compilation time varies across benchmarks, particularly for those with higher gate density.
We also observe spikes in compilation time, which can be primarily attributed to the characteristics of our algorithm.
Locally optimal scheduling decisions can lead to an expanded search space in subsequent steps, resulting in fluctuations in compilation time.

\subsection{Compilation Time and Fidelity Trade-off}

\begin{figure}[h]
    \centering
    \includegraphics[width=1\linewidth]{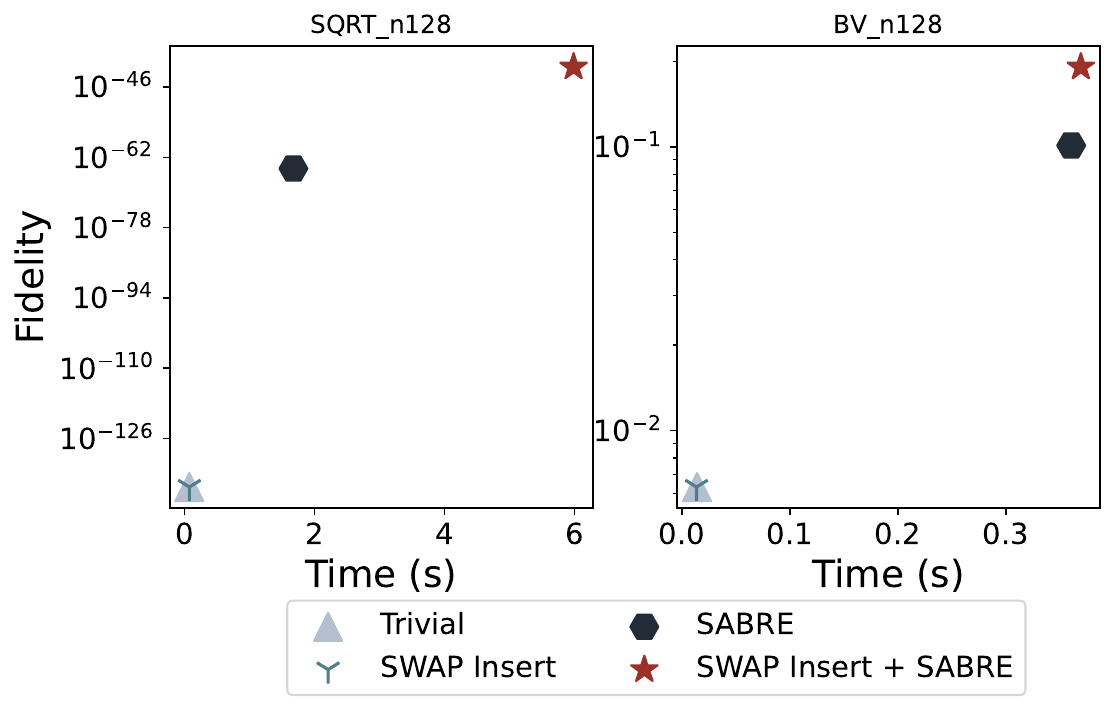}
    \caption{Comparison of the compilation time and fidelity.}
    \label{fig:fidelity_time_analysis}
\end{figure}

Furthermore, we conduct analysis on the compilation time of different techniques with respect to the final fidelity within the MUSS-TI. Fig.~\ref{fig:fidelity_time_analysis} presents a comparison of fidelity versus compilation time for different compilation strategies within the MUSS TI framework. We select two candidate applications such as SQRT and BV with qubit size 128, which stands for the complex and simple application respectively. In Fig.~\ref{fig:fidelity_time_analysis} left and right subplots correspond to the results of these two benchmark circuits. In both cases, the combined strategy (\texttt{SWAP Insert + SABRE}) achieves the highest fidelity, though at the cost of increased compilation time. 
These results demonstrate the effectiveness of combining mapping and gate insertion techniques in maximizing fidelity, while also highlighting the common trade-off between compilation time and fidelity.

\subsection{Multiple Entanglement Zone Analysis}

\begin{figure}[h]
    \centering
    \includegraphics[width=0.8\linewidth]{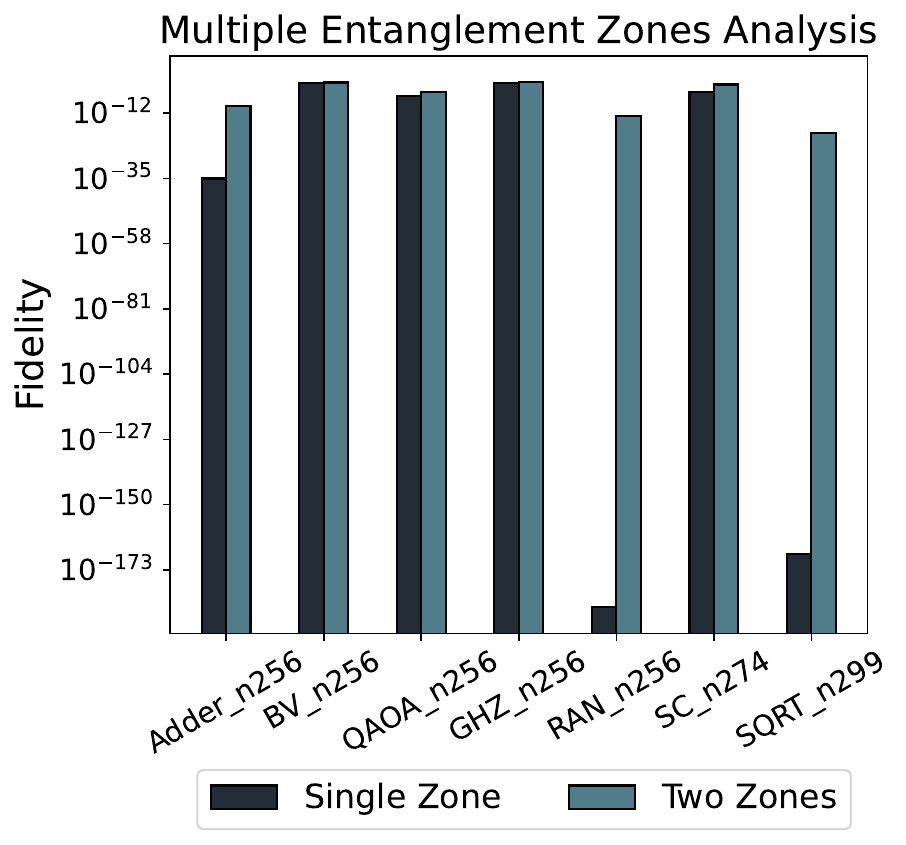}
    \caption{ Analysis of multiple entanglement zones. We compare the circuit fidelity when using one entanglement zone versus two.}
    \label{fig:multiple_zone_analysis}
\end{figure}

Currently, we consider only a single entanglement zone, where all two-qubit gates must be shuttled for entangling operations. This constraint requires certain qubits to be moved into the zone, which can generate significant heat in the trap and negatively impact the final fidelity. To consider this issue, we analyze the impact of introducing multiple entanglement zones, specifically comparing the cases of one versus two zones. The results are shown in Fig.~\ref{fig:multiple_zone_analysis}, where we evaluate all larger applications with sizes ranging from 256 to 299 qubits. Across most applications, the two entanglement zones configuration yields higher fidelity, highlighting the benefits of distributing entangling operations across multiple zones. This demonstrates the potential of multi-zone architectures in mitigating heat accumulation and improving overall circuit performance.

\begin{figure*}[h]
    \centering
    \includegraphics[width=1\linewidth]{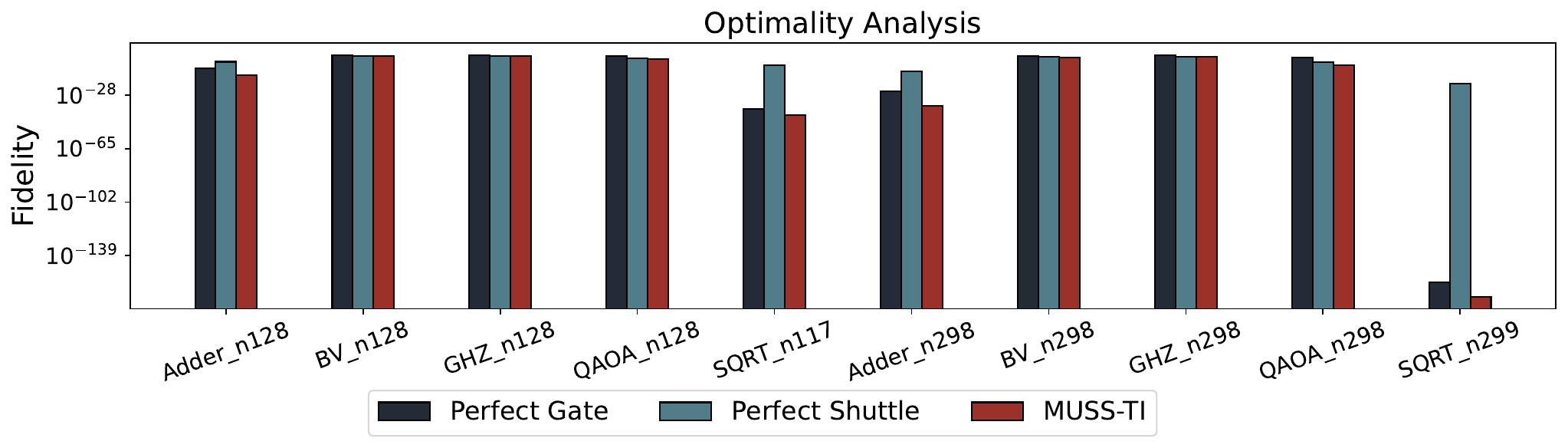}
    \caption{
    Optimality Analysis of MUSS-TI. We consider two idealized scenarios: the \texttt{perfect gate} and \texttt{perfect shuttle} cases. The \texttt{perfect gate} scenario assumes a high-fidelity gate model, while the \texttt{perfect shuttle} scenario assumes that shuttling operations introduce no heat into the trap.
    }    
    \label{fig:optimality_analysis}
\end{figure*}

\subsection{Optimality Analysis}
\label{sec: optiamlity analysis}

In this section, we discuss the level of fidelity that MUSS-TI can achieve under ideal conditions. Given the computational difficulty of finding optimal mapping and routing solutions due to the exponentially growing search space, we introduce an idealized setting to assess the potential upper bound of performance. Specifically, we examine two ideal scenarios: one with no heat introduced by shuttling (perfect shuttling), and another with exceptionally high two-qubit gate fidelity, set to 0.9999 (perfect gates). The results are shown in Fig.~\ref{fig:optimality_analysis} and demonstrate that MUSS-TI achieves fidelity close to that of the perfect gate and perfect shuttle conditions in most applications.
In most cases, the perfect gate provides greater fidelity improvement than the perfect shuttle. This indicates that MUSS-TI will prioritize the use of on-chip gate, and it can also perform well when off-chip worse than on-chip.

The experimental results also reveal that the factors influencing circuit fidelity vary significantly across different applications. Although it might seem intuitive that circuits with more gates would be more sensitive to gate fidelity, our findings indicate the opposite. Specifically, circuits with more gates tend to benefit less from improvements in gate fidelity and more from reductions in shuttle-induced errors. This is largely due to the increased number of shuttling operations required by such circuits, which can introduce thermal effects that degrade the fidelity of subsequent gate executions. Consequently, in most cases, circuits with a higher gate count experience more pronounced fidelity degradation as a result of shuttling.

\section{Related Work}

\textbf{Single-trap Trapped-ion Compilers.} The single-trap trapped ion have been first attract to seeking the optimization opportunity. There are many methods such as integer linear programming and heuristic functions that are specifically tailored for single-trap trapped-ion devices~\cite{wu2019ilp_iontrap, wu2021tilt_iontrap, stevens2017automating, wu2024boss}. However, scalability remains a challenge for single trapped-ion devices and IonQ revealed their capacity to manage up to 160 qubits within their architecture \cite{ionq2018harnesses}. Furthermore, the shuttling operations in these devices differ from those used in EML-QCCD considered in this work.

\textbf{QCCD Compilers.} Building on the foundational work that introduced QCCD compilers and explored microarchitecture and architectural simulation for predicting the performance of near-term QCCD systems~\cite{murali2020architecting_iontrap}, several subsequent studies have focused on improving shuttle scheduling. For example, heuristic approaches have been proposed for optimizing linearly connected QCCD devices~\cite{saki2022muzzle, upadhyay2022shuttle}, Boolean satisfiability has been used to obtain exact solutions~\cite{schoenberger2024using}, parallelism has been enhanced in~\cite{ovide2025exploring}, and efficiency has been improved in~\cite{bach2025efficient, zhu2025s}. However, these works primarily focus on near-term QCCD devices, whereas our work targets next-generation devices consisting of thousands of qubits, whose feasibility has already been demonstrated experimentally. Moreover, prior approaches do not consider the entanglement zone, which introduces an additional physical constraint addressed in MUSS-TI.

\textbf{QCCD experimental demonstration.} Number of experimental studies utilizing trapped-ion devices have been executed, demonstrating their formidable capability in executing quantum applications. Notably, Quantinuum has introduced the QCCD device~\cite{moses2023race} and conduct several experiments on ~\cite{iqbal2024non, he2023alignment, Liu_2025} demonstrating its power in near-term applications. 
Recent studies have advanced the scalability of quantum systems by interconnecting multiple QCCDs via optical fibers~\cite{kwon2024multi, akhtar2023high}. Integrated optical waveguides have enabled simultaneous control of QCCD zones on a single chip~\cite{mordini2025multizone}, while Mach-Zehnder modulators have facilitated optical resource sharing across separate chips~\cite{hogle2023high}. Additionally, entanglement between ion trap qubits over 200 meters has been experimentally demonstrated~\cite{krutyanskiy2023entanglement}, showcasing the feasibility of long-distance quantum links and paving the way for entanglement-connected QCCD architectures. In line with this experimental progress, our work aims to prepare for upcoming large-scale trapped-ion systems and has the potential to inform the microarchitecture design of such systems.

\textbf{Mapping and Routing.} Many mapping and routing problems have been studied to translate high-level quantum applications or circuits into hardware-compatible forms through various compilers~\cite{zhu2025quantum}. Among the most extensively investigated platforms are superconducting devices, with research spanning from early work on general mapping and routing~\cite{baker2020time, liu2021relaxed, liu2022not, lye2015determining, murali2019noise, nannicini2022optimal, oddi2018greedy, patel2022geyser_atom, smith2021error, das2021adapt, wille2014optimal, aleksandrowicz2019qiskit, sivarajah2020t}, circuit synthesis~\cite{das2021adapt, das2021jigsaw, patel2021qraft, patel2020disq, smith2021error, stein2023q} and pulse-level optimization~\cite{gokhale2020optimized, liang2022pan, meirom2023pansatz, shi2019optimized, zhu2024leveraging}, to more recent application-specific compilers targeting quantum error mitigation~\cite{das2021adapt, das2021jigsaw, patel2021qraft, patel2020disq, smith2021error, stein2023q}, quantum error correction~\cite{wang2024optimizing, yin2025qecc, yin2024surf}, and Hamiltonian simulation~\cite{cowtan2019phase, paykin2023pcoast, meijer2023towards, li2021paulihedral, jin2024tetris, de2024faster, lao20222qan, alam2020efficient, alam2020circuit, alam2020noise}.
Aside from superconducting systems, neutral-atom devices have also been a key focus in managing atom movements~\cite{lin2024reuse, wang2024atomique, bochenneutralatom, hanruiwangQpilot, tan2022qubit}. While they share a similar goal of minimizing the overhead in transforming circuits into hardware-compatible forms, the underlying hardware constraints differ significantly and thus cannot be directly applied to our work.

\section{Conclusion}

Towards practical, large-scale quantum computing, entanglement-module-linked trapped-ion systems offer a promising approach, leveraging efficient quantum information transfer via fiber entanglement. In this work, we propose MUSS-TI, a framework specifically designed for this device. Our results demonstrate that MUSS-TI is well-suited for large-scale applications, offering reduced shuttling operations and improved overall performance.

\textbf{Outlook.} To support large-scale quantum applications, it is essential to implement quantum error correction protocols~\cite{wang2024optimizing, yin2025qecc, yin2024surf}. Exploring how to tailor these codes to EML-QCCD architectures remains a crucial step toward achieving fault-tolerant quantum computation. Additionally, enhancing compiler efficiency and effectiveness is vital for bridging the gap between high-level algorithms and the constraints of underlying hardware.

\begin{acks}
X. Wu and C. Zhu contributed equally to this work. This work was partially supported by the National Key R\&D Program of China (Grant No.~2024YFB4504004), the National Natural Science Foundation of China (Grant. No.~12447107), the Guangdong Provincial Quantum Science Strategic Initiative (Grant No.~GDZX2403008, GDZX2403001), the Guangdong Provincial Key Lab of Integrated Communication, Sensing and Computation for Ubiquitous Internet of Things (Grant No. 2023B1212010007), the Quantum Science Center of Guangdong-Hong Kong-Macao Greater Bay Area, and the Education Bureau of Guangzhou Municipality.
\end{acks}


\newpage
\bibliographystyle{ACM-Reference-Format}
\bibliography{sample-base}

\end{document}